\documentclass{amsart}

\usepackage{t1enc,newlfont}
\usepackage[latin5]{inputenc}
\usepackage[dvips]{graphicx}

\begin{document}
\title[Bond Distortions in Armchair Type Single Wall Carbon Nanotubes]
{Bond Distortions in Armchair Type Single Wall Carbon Nanotubes}

\author{N. Sunel$^1$, E. Rizaoglu$^2$, K. Harigaya$^3$ and O. Ozsoy$^1$}

\address{$^1$Department of Physics, Faculty of Arts and Sciences, Gaziosmanpasa University, 60240 Tokat, Turkey.\\ $^2$Department of Physics, Faculty of Sciences, Istanbul University, 34459 Istanbul, Turkey.\\ $^3$ Nanotechnology Research Institute, AIST, Tsukuba 305-8568, Japan.}

\email{ozsoyo@gop.edu.tr}


\begin{abstract}

The energy band gap structure and stability of (3,3) and (10,10)
nanotubes have been comparatively investigated in the frameworks
of the traditional form of the Su-Schrieffer-Heeger (SSH) model
and a toy model including the contributions of bonds of different
types to the SSH Hamiltonian differently. Both models give the
same energy band gap structure but bond length distortions in
different characters for the nanotubes.

\end{abstract}

\maketitle

\section{Introduction}

A single-wall carbon nanotube (SWCNT) is an empty tube of graphene
consisting of hexagonally arranged carbon atoms. In graphene,
there are two different rim shapes, armchair and zigzag. For an
armchair SWCNT, the hexagon rows are parallel to the tube axis.
The $\pi$-electronic structure of an armchair SWCNT arises from
the $\pi$-structure of graphene. Each carbon atom in graphene
contributes to the structure with one electron in the
$2p_{\mathrm{z}}$ orbital perpendicular to the plane of the sheet.
Generally, the overlap of $\pi$-orbitals due to the curvature in
nanotubes are neglected for moderate curvatures. If $n$ is the
number of two-carbon sites (dimers), {\it i.e.} the nearest
neighbors on polyacetylene (PA) chain which is the prototype
polymer of graphene, the nanotube is labeled as ($n$,$n$). One of
the most important properties of armchair nanotubes is that they
show metallic behavior [1].

In the present treatise the tight-binding approximation, which is
sometimes known as the method of linear combination of atomic
orbitals, is used. This approximation deals with the case in which
the overlap of atomic wave functions is enough to require
corrections to the picture of isolated atoms but not so much as to
render the atomic description completely irrelevant. It is mostly
used for describing the energy bands arising from the partially
filled d-shells of transition metal atoms and for describing the
electronic structure of insulators. Moreover, the tight-binding
approximation provides an instructive way of viewing Bloch levels
complementary to that of the nearly free electron picture,
permitting a reconciliation between the apparently contradictory
features of localized atomic levels on the one hand and free
electron-like plane-wave levels on the other [2].

The tight-binding approximation was originally developed by
Su-Schrieffer-Heeger (SSH) [3] for conducting polymers (1D
systems) and then extended to two-dimen\-sional systems by
Harigaya [4, 5]. Harigaya's model preserves the fixed-length
constraint of one-dimensional polymer chain and hence it contains
a single Lagrange multiplier. In most applications of this model
to graphene and to tubes constructed from graphene, the constraint
has still been used in the same form, that is all bond distortions
are summed without considering the type of bonds and this sum is
assumed to vanish. However, two different types of bonds appear in
graphene and so in tubes, tilt and right (see Fig. 1). It would
also be worthwhile to point out that bond length difference of
hexagon structure have been reported by the calculations on
graphene and nanotubes [5, 6].

\begin{figure}[ht]
\begin{center}
\includegraphics[angle=0, width=9cm]{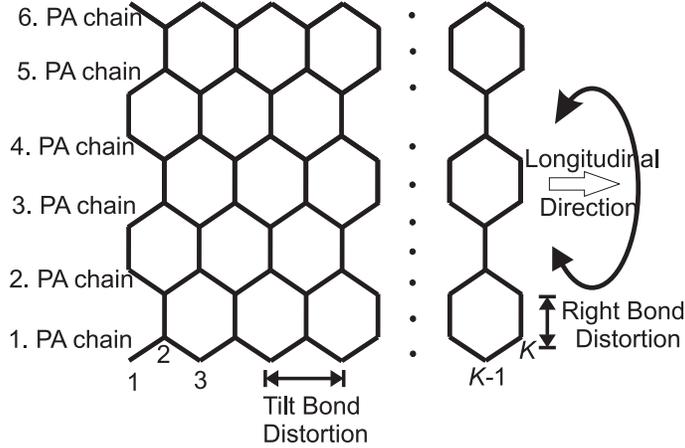}
\caption{Inter-chain coupling structure of a (3,3) armchair type
nanotube and bond distortions. }
\end{center}
\end{figure}

Considering this fact, in this work on armchair type nanotubes, we
present for the first time the modification in Harigaya's model by
taking the contributions of bonds of different types to the SSH
Hamiltonian differently. This automatically leads us to separate
the constraint into two constraints, vanishing of the sum of right
bond distortions and vanishing of the sum of tilt bond
distortions. In this way we build a toy model which provides more
freedom for lattice relaxations. We have already mentioned the
very preliminary results of this toy model in our work in [7]. In
our second work [8], we evaluated the electronic band structure of
(3,0) nanotube with periodic boundaries in the framework of this
toy model and in the Harigaya's model comparatively. We observed
that the tiny energy gap appearing in Harigaya's model was lost
when our toy model has been used. This result consists with the
fact that zigzag nanotubes ($n$,0) are metallic when $n$ is any
multiple of 3, and semiconducting when $n$ cannot be divided by 3.
This is determined by whether the $K$ and $K^{\prime}$ points of
the graphite meet with the one-dimensional Brilloune zones
determined by the geometry or not.

The ($n$,$n$) armchair nanotubes are always metallic for all the
integers $n$ and metallic behavior of (3, 3) armchair nanotube has
also experimentally being shown [9, 10]. Recently, Li {\it et al.}
[11] have grown free-standing SWCNTs. Their diameter is as small
as 0.4 nm. The (3,3) armchair nanotubes are among the possible
structures of this size [11]. This is why we deal with here with
(3,3) armchair nanotube in the framework of our toy model. On the
other hand, the commonly observed diameter of SWCNT by experiments
is known as about 1.4 nm which corresponds to that of (10,10)
SWCNT. Therefore, we also test our toy model with this larger
diameter nanotube.

\section{MODEL}

The SSH model Hamiltonian
\begin{equation}
H_{\mathrm{SSH}}=-\sum_{\langle i,j \rangle,\sigma}\Big[t_0 -
\alpha
(u_{i}^{(j)}-u_{j}^{(i)})\Big](c^{\dagger}_{i,\sigma}c_{j,\sigma}
+\mathrm{h.c.})
\end{equation}
\begin{equation*}
+\frac{\kappa}{2}\sum_{\langle i,j
\rangle}\Big[(u_i^{(j)}-u_{j}^{(i)} - C)^2-(C)^2\Big]\,,
\end{equation*}
which had originally been written for 1D systems, was directly
applied to 2D systems without any modification by Harigaya [5].
Here, $\langle i,j \rangle$ is the nearest-neighbor carbon-carbon
atom pairs and $t_0$ is the hopping integral of the undimerized
system. The second term represents the dimerization due to
$\sigma$ skeleton with free involving $\pi$-electrons. $\alpha$ is
the electron-lattice coupling constant, $\kappa$ is the effective
spring constant. The operator $c_{i,\sigma}^{\dagger}$
($c_{i,\sigma}$) creates (annihilates) a $\pi$-electron at the
$i$-th carbon atom with spin $\sigma$. $u_i^{(j)}$ is the
displacement of the $i$-th atom along the $j$-th one, whereas in
the original SSH model Hamiltonian $u_i^{(j)}$ is perpendicular to
carbon-hydrogen bond direction for both trans-PA and cis-PA. The
term $u_i^{(j)} - u_j^{(i)}$, when considered for trans-PA or
cis-PA, denotes the projection of the bond length difference along
chain axis. But for nanographite, the same term denotes directly
the bond length difference. $C$, in the last term, is the Lagrange
multiplier in the self-consistent method which has been inserted
in the original SSH Hamiltonian due to the fixed length constraint
when it was firstly used for trans-PA and cis-PA [12]. Harigaya
preserved the same constraint for 2D systems [4, 5]. This
constraint binds the distortions in both the length and
circumference directions of the tube and, in a sense, relatively
restricts the lattice relaxation.

In trans-PA and cis-PA, all the bonds are of the same type. An
armchair (and also a zigzag) SWCNT can be thought of to be built
by parallel trans-PA with inter-chain coupling as seen in Fig.\,1.
Therefore in nanotubes there are two different types of bonds,
tilt and right bonds.

During the lattice relaxation, of course, the relative distances
between the carbon atoms in hexagons and the angles in hexagons
become different. In the toy model, which we wish to built in the
present work, as a first approximation we are going to neglect the
changes in the angles in hexagons and keep only the changes in the
relative distances. Furthermore, we would like to modify the SSH
model Hamiltonian by taking the contributions of the two different
bond types and rewrite the SSH model Hamiltonian as follows:

\begin{eqnarray}
  H_{\mathrm{SSH}}& = & -\sum_{{\langle i,j \rangle _\mathrm{t},\sigma}}\Big[t_0^{\mathrm{t}}-\alpha^{\mathrm{t}}(u_{i}^{(j)}-u_{j}^{(i)})\Big](c^{\dagger}_{i,\sigma}c_{j,\sigma}+\mathrm{h.c.}) \\ \nonumber
                  &   &-\sum_{{\langle i,j \rangle _\mathrm{r},\sigma}}\Big[t_0^{\mathrm{r}}-\alpha^{\mathrm{r}}(u_{i}^{(j)}-u_{j}^{(i)})\Big](c^{\dagger}_{i,\sigma}c_{j,\sigma}+\mathrm{h.c.})   \nonumber \\
                  &   & +\frac{\kappa^{\mathrm{t}}}{2}\sum_{\langle i,j \rangle _\mathrm{t}}\Big[(u_i^{(j)}-u_{j}^{(i)}-C^{\mathrm{t}})^2-(C^{\mathrm{t}})^2\Big] \nonumber \\
                  &   & +\frac{\kappa^{\mathrm{r}}}{2}\sum_{\langle i,j \rangle _\mathrm{r}}\Big[(u_i^{(j)}-u_{j}^{(i)} -C^{\mathrm{r}})^2-(C^{\mathrm{r}})^2\Big]\,, \nonumber
\end{eqnarray}
where $\langle i,j \rangle _\mathrm{t}$ and $\langle i,j \rangle
_\mathrm{r}$ denote tilt and right bonds, respectively. With this
separation it seemed to us natural to use two different Lagrange
multipliers, $C^{\mathrm{t}}$ and $C^{\mathrm{r}}$. To be able to
insert these multipliers in the Hamiltonian we have to consider
two constraints: vanishing separately the sum of all tilt bond
distortions and the sum of all right bond distortions, {\it i.e.}
$\sum v_{i,j}^{\mathrm{t}} = 0$ and $\sum v_{i,j}^{\mathrm{r}} =
0$, where $v_{i,j} \equiv \alpha(u_i^{(j)}-u_{j}^{(i)})$. This
means that lattice relaxations may gain more freedom. The mean
value of the sum in the first constraint is obviously proportional
to the length of tube and the mean value of the sum in the second
constraint is related to the circumference of tube. As a matter of
fact, in the literature Ono and Hamano considered also two
constraints during the study of Peierls distortions in a
two-dimensional electron-lattice system described by SSH type
model [13].

In this case the total energy of the system reads as
\begin{equation}
E_{\mathrm {T}} = \sum_{i, \sigma}^{\prime} \varepsilon_{i,
\sigma} + \frac{1}{2\gamma^{\mathrm{t}}}\,\sum_i
[v_i^{\mathrm{t}}]^2 + \frac{1}{2\gamma^{\mathrm{r}}}\,\sum_i
[v_i^{\mathrm{r}}]^2\,,
\end{equation}
where $\varepsilon_{i, \sigma}$ are the eigenvalues of Eq.\,(2).
The self-consistent equation for the lattice is

\begin{eqnarray}
v_i^{{\mathrm{t,r}}}=2\gamma^{{\mathrm{t,r}}}\Big[\frac{\alpha^{{\mathrm{t,r}}}
C^{{\mathrm{t,r}}}}{\gamma^{{\mathrm{t,r}}}}
-\frac{1}{N_b^{{\mathrm{t,r}}}}\frac{\gamma^{{\mathrm{t,r}}}}{\alpha^{{\mathrm{t,r}}}}\sum_{j,\sigma}^{\prime}B^{\dag}_{i+
1,j,\sigma}B_{i,j,\sigma}\Big]\,.
\end{eqnarray}
where ${\mathbf{B}}$'s are the eigenvectors of Eq.\,(2) and
$\gamma^{\mathrm{t,r}}=(\alpha^{\mathrm{t,r}})^2/\kappa^{\mathrm{t,r}}$.
The prime means that the sum is over the filled states, the number
of which are equal to half of the total number of carbon sites.
$N_{\mathrm b}^{{\mathrm{t,r}}}$'s are the total number of tilt
and right $\pi$-bonds, respectively. Eqs.\,(2)-(4) are solved
numerically by iteration method [7, 8].

\section{RESULTS AND DISCUSSIONS}

We describe the geometry of nanotubes with bonds. During the
numerical evaluations we considered the diameter and lengths of
(3,3) tubes as 0.5 nm and 0.73-36.36 nm, respectively since
experimentally, 0.4 nm-sized carbon nanotubes have been reported
to exist [9-11] and also 0.33 nm-sized carbon nanotubes were also
grown from a larger SWCNT inside an electron microscope [14]. For
the purpose of checking our numerical evaluations, we have
calculated the diameters of (3,3) and (5,0) nanotubes. Although we
could find the given diameter for (5,0) nanotube we could not find
the given diameters 0.4 nm and 0.33 nm for (3,3) nanotube, instead
we found 0.5 nm, when both calculations had been done with the
same parameters derived from the experimental work [11]. Moreover,
we repeat our numerical calculations for (10,10) nanotube which
has the diameter 1.4 nm, the commonly observed diameter of SWCNT
by experiments, when the tube length varies between 0.73 nm and
16.0 nm.

The calculations on (3,3) armchair type nanotube is considered
with $N = 2n$ and the PA chain length $K$ varying between 6 (0.73
nm) and 300 (36.36 nm), that is with maximum 1800 C-atoms. For the
(3,3) armchair type open ended nanotube, the number of right bonds
is 900 and the number of tilt bonds is 1794 for $K = 300$. In the
periodic boundaries case, the number of right bonds keeps itself
while the number of tilt bonds increases 6 more.

The calculations on (10,10) armchair type nanotube is considered
with $N = 2n$ and $K$ varying between 6 (0.73 nm) and 132 (16.0
nm), that is with maximum 2640 C-atoms. For the (10,10) armchair
type open ended nanotube, the number of right bonds is 1320 and
the number of tilt bonds is 2620 for $K = 132$. In the periodic
boundaries case, the number of right bonds keeps itself while the
number of tilt bonds increases 20 more.

Firstly, we studied the electronic band structure as (3,3)
armchair nanotube evolves from carbon sheet, in the cases the tube
is open ended and possesses periodic boundaries, in the framework
of our toy model and in that of Harigaya's model comparatively by
taking $t_0$ and $\alpha$ values the same for tilt and right
bonds. We realized this by multiplying each one of the hopping
integrals responsible for sheet, open ended tube and tube with
periodic boundaries by parameters denoted by $\beta_{\mathrm S}$,
$\beta_{\mathrm N}$ and $\beta_{\mathrm T}$, respectively and by
varying these parameters from zero to one; that is by introducing
gradually the interactions characterizing the aforementioned
structures. In this way we can show continuously the evolution of
electronic band structure as geometry transforms. In both models,
we obtained identical electronic band gap structures shown in
Fig.\,2. Hence, contrary to (3,0) zigzag nanotube [8], there is
not any difference between the electronic band structures of
metallic (3,3) armchair nanotube regarding both models. We can
explain this fact in the following way: The electronic band
structure in the scale of the total $\pi$-electron energy bands
seem similar, because the bond alternation amplitude is one order
of magnitude smaller than that of PA (see Figs.\,3. and 4) and the
magnitude of $u$ is about 0.03 \AA \,\,in trans-PA [3]. The
electronic band structure of (10,10) nanotubes gives no
significant change in comparison with the electronic band
structure given in Fig.\,2 for (3,3). Thus we are not going to
plot it separately.

\begin{figure}[ht]
\begin{center}
\includegraphics[angle=0, width=9cm]{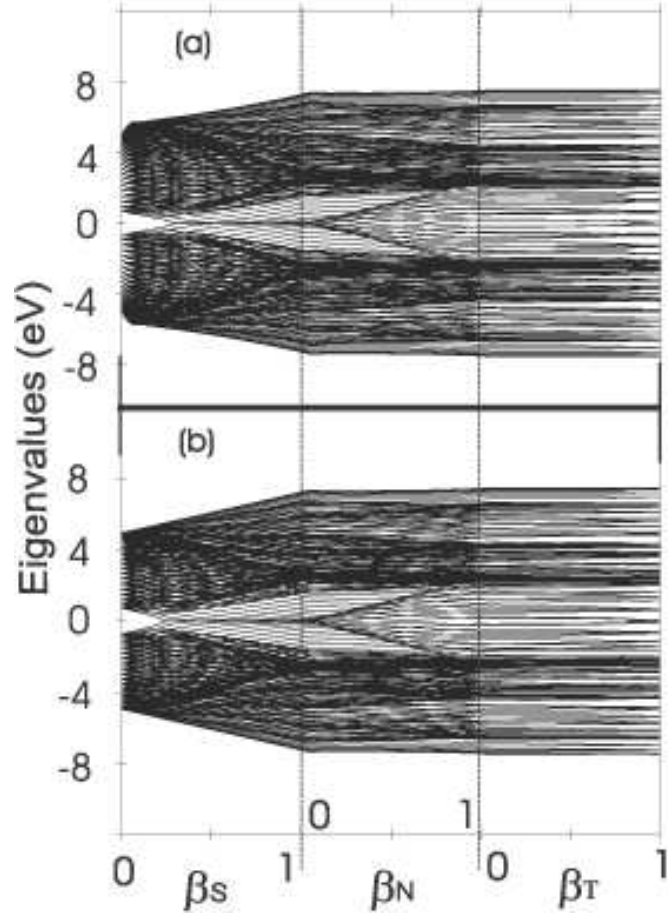}
\caption{Electronic band structure evolutions from carbon sheet to
SWCNT for (3,3) armchair type nanotube with $K = 42$ for models
including (a) one constraint and (b) two constraints.
$\beta_{\mathrm{S}}$, $\beta_{\mathrm{N}}$ and
$\beta_{\mathrm{T}}$ represent the evolution parameters for carbon
sheet, open ended nanotube and nanotube with periodic boundaries.
No band gap appears as expected. For (10,10) nanotube the same
electronic band structure was obtained. Only, the marked regions
appeared darker. This is because of the jump of the number of
energy eigen-values from 504 to 840 when (10,10) nanotube is
considered instead of (3,3) for the same $K$ value, $K = 42$.}
\end{center}
\end{figure}

Secondly, we consider the stability problem of (3,3) and (10,10)
armchair nanotubes in the framework of both models when nanotubes
possess periodic boundaries.

According to the Harigaya's model, we numerically calculated the
bond distortions via the simplified form of Eq.\,(4). For (3,3)
nanotube the results can be expressed as follows (Fig.\,3):  The
right bond distortions are twice of the tilt bond distortions. The
right bond distortions are always negative, that is these bonds
shrink while the tilt bond distortions are always positive, that
is tilt bonds stretch. Besides, at the beginning the distortions
increase up to $K = 12$ and then tend to decrease. For $K = 16$
shrinking of right bonds suddenly turns to stretching and, at the
same time, stretching of tilt bonds suddenly turns to shrinking.
For $K=22$ there is no bond distortions for both types of bonds.
Therefore, the first thing to be noted is that in order to be able
to obtain reasonable results in agreement with the literature one
must consider the tubes longer than 2.67 nm ($K=22$)[6]. After
this special $K$ value (for tubes longer than 2.67 nm), the tube
length-oscillations of tilt bond distortions show decaying. The
half period of oscillations about some mean tilt bond distortion
values is 3. These mean values at first decay and then remain
almost the same after $K=212$. This means that, as the length of
tube becomes longer, the tilt bond stretching values repeat a
decaying increasing-decreasing behavior about almost constant bond
distortions. Furthermore, the amplitude of oscillation decays,
that is for longer tubes oscillation of tilt bond distortions
vanish and after a certain large $K$ value (approximately 400) the
bond distortions approach a definite nonzero value. Meanwhile, the
right bond shrinkages behave similarly and they also tend to
nonzero values. These are unexpected results because it is known
that long enough tubes are stable.

\begin{figure}[ht]
\begin{center}
\includegraphics[angle=0, width=10cm]{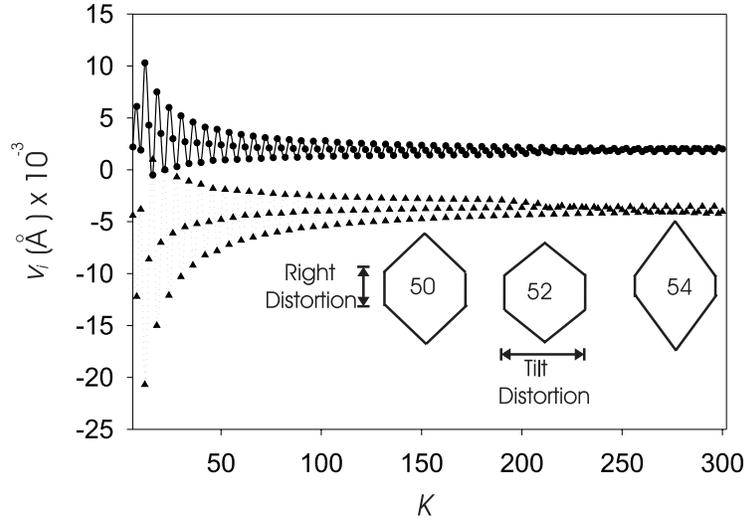}
\caption{The filled circles and the filled triangles represent the
tilt and right bond distortions, respectively when Harigaya's
model is used. $K$ is the number of sites in trans-PA chains. In
the inset the forms of hexagons for $K = 50, 52$ and $54$ are
depicted. Since the right bonds always shrink (the amount of
shrinkage is maximum for $K = 54$ and minimum for $K = 52$) and
the tilt bonds always stretch (the amount of stretching is minimum
for $K = 52$ and maximum for $K = 54$), the diameter of the tube
decreases while it's length increases.}
\end{center}
\end{figure}

As for (10,10) nanotube, a repetition of the above evaluations now
gives the result summarized in Fig.4(a). There is not any
systematic change in the tilt and right bond distortions as $K$
increases. Both of them decay immediately to zero. Hence the tube
is stable except for very small $K$ values. This is quite the
expected result because wide enough tubes might be stable for
smaller $K$ values that correspond to shorter tubes. Therefore the
shorter tubes would be necessary in order to observe bond
alternations otherwise the longer tubes will show relatively small
alternations.

\begin{figure}[ht]
\begin{center}
\includegraphics[angle=0, width=11cm]{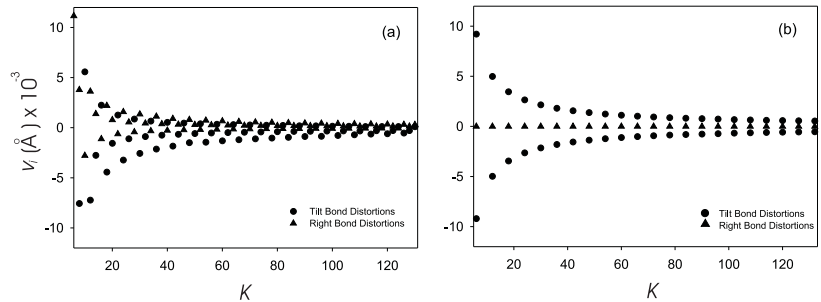}
\caption{(a) The filled circles and the filled triangles denote
the tilt and right bond distortions, respectively when Harigaya's
model is used, for (10,10) nanotube. (b) The bond alternations in
the case the toy model is used.}
\end{center}
\end{figure}

According to our toy model, we consider Eq.\,(4) for the bond
distortions by taking again $t_0$ and $\alpha$ values the same for
tilt and right bonds. For both (3,3) and (10,10) nanotubes, all
the right bond distortions vanish while tilt bond distortions
appear as bond alternations when $K$ is divisible by 3. For both
(3,3) and (10,10) tubes, the tilt bond distortions vanish also
when $K$ is not divisible by 3. Of course, vanishing of tilt bond
distortions for $K$ values not divisible by 3 is a handicap for
the toy model.

The bond distortions for (3,3) armchair type SWCNT with periodic
boundaries are illustrated in Fig.\,4. As seen from this figure,
equal amount of stretching and shrinkage of alternate bonds occur.
The tilt bond alternations start with $\pm 0.0306$ \AA\, for $K =
6$ and drastically fall to $\pm 0.0066$ \AA\, at $K = 36$. In
other words, the lattice structure relaxation starts at $K = 36$.
It is clear that tilt bond alternation values decrease
exponentially for small $K$ values while they decrease
monotonically for large $K$ values. Although we did the
calculations up to $K = 400$, we terminated the graphic at $K =
300$ since we observed the continuation of monotonic decreasing of
tilt bond alternations. At this point, we believe that the
explanation of the aforementioned difference in the bond
distortions for various $K$ values (multiple of 3 or not) is
vital.

\begin{figure}[ht]
\begin{center}
\includegraphics[angle=0, width=11cm]{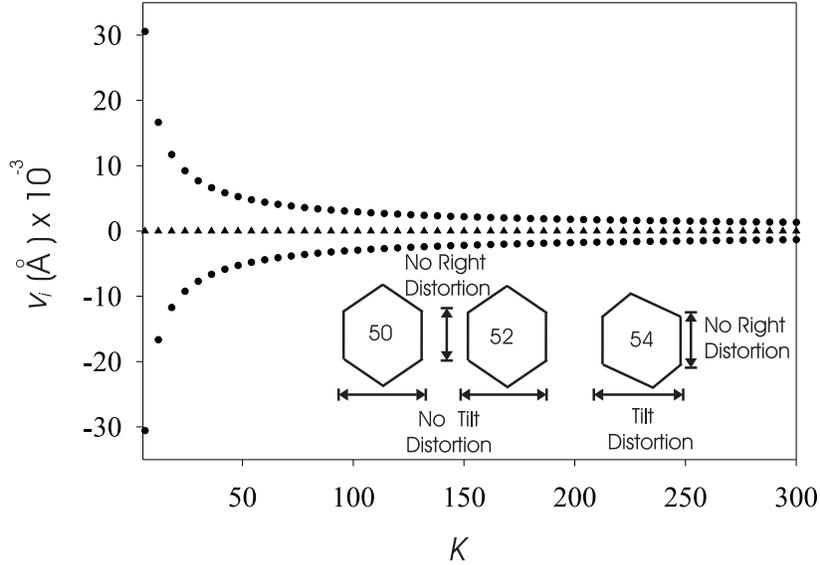}
\caption{The filled circles and the filled triangles denote the
tilt and right bond alternations, respectively when our toy model
is used, for $K$ divisible by 3. In the inset the forms of
hexagons for $K = 50,\, 52$ and $54$ are depicted. Neither the
diameter nor the length of tube changes.}
\end{center}
\end{figure}

The bond distortions for (10,10) armchair type SWCNT with periodic
boundaries are illustrated in Fig.\,4(b). We have the same bond
distortion behavior, bond alternations.

To see the strength of dimerization, we plot in Fig.6 the $1/K$
variation of tilt and right bonds and also the average of the
absolute values of bond alternations for both (3,3) and (10,10)
nanotubes when $K$ is divisible by 3. $\langle |v_i| \rangle$
decreases linearly for both tubes. The extrapolated value at $K
\to \infty $ is 0.0005 nm [5,15] for (3,3) nanotube  (Fig.6(c)).
For (10,10) nanotube, $\langle |v_i| \rangle$ reaches to zero for
$K \approx 10^4$  (Fig.6(d)). Moreover, from the $1/K$ variation
of the tilt and right bonds one sees the length differences
between the long and short bonds to be 0.002 nm and 0 nm,
respectively for the (3,3) and (10,10) nanotubes (Fig.6 (a) and
(b)).

To bring to a conclusion, although for (3,3) armchair type SWCNTs
with periodic boundaries the two models give different stability
properties, (10,10) SWCNTs are stable according to both models.
But still there is a difference. The toy model puts forth bond
alternations. When the size of the bond alternations of (3,3) and
(10,10) SWCNTs are compared,  a decrease is observed. For
instance, for $K = 6$ the tilt bond alternations is $\pm 0.0306$
\AA\ for (3,3) and $\pm 0.0009$ \AA\ for (10,10) and for $K = 132$
the tilt bond alternations is $\pm 0.0024$ \AA\ for (3,3) and $\pm
0.0005$ \AA\ for (10,10). Hence, the bond alternations diminishes
at $66\%$ for $K = 132$. Therefore, the tube diameter strongly
affects the bond alternation amplitudes.

\begin{figure}[ht]
\begin{center}
\includegraphics[angle=0, width=12cm]{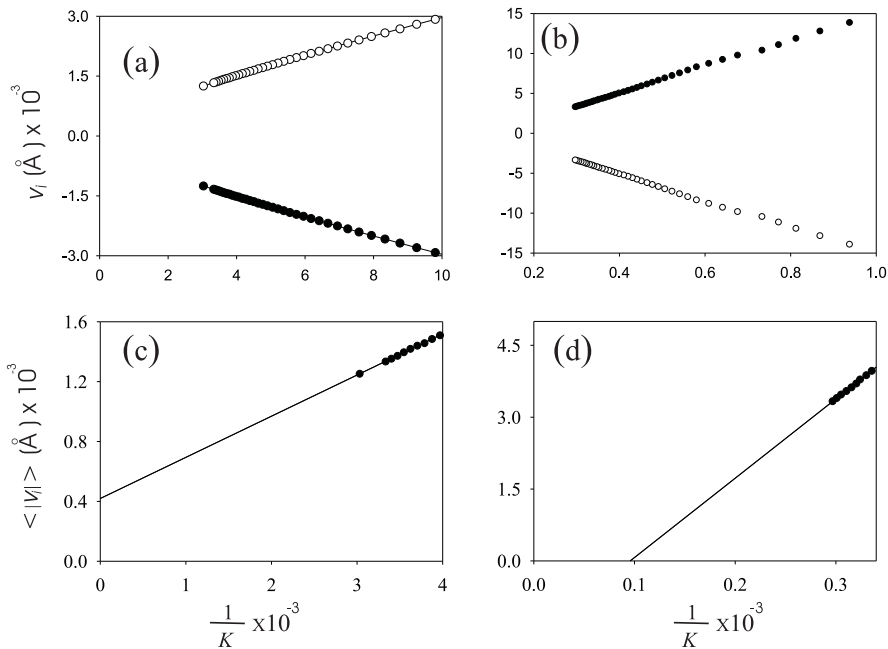}
\caption{The $1/K$ variation of bond variables. $\langle |v_i|
\rangle$'s for the (3,3) and (10,10) nanotubes are shown in (a)
and (c), respectively- whereas. (b) and (d) show the $v_i^{\mathrm
r, t}$.}
\end{center}
\end{figure}

\section{CONCLUSION}

In the progress of taking the coefficients of the contributions of
the tilt and right bond distortions to the SSH model Hamiltonian
differently and following Stafstrom {\it et el.}'s work on the
inclusion of Lagrange multiplier in the Hamiltonian [12], we have
to use two constraints, one for the vanishing of the sum of right
bond distortions and the other for the vanishing of the sum of
tilt bond distortions. Here, a toy model obtained in this way is
applied to calculate the energy band gap and bond variations with
the number of rows, $K$, of (3,3) and (10,10) armchair type
nanotubes of diameters 0.5 nm and 1.4 nm, respectively. There is
not any change in the energy band gap structure in the large
energy scale. However contrary to this, bond length alternations,
which were absent in the one constraint Harigaya's model, have
been observed when $K$ is divisible by 3. These alternations tend
to vanish as the $K$ values increase, that is for long enough
tubes. Unfortunately, the vanishing of bond length distortions
when $K$ is not divisible by 3 would be an artifact of the model.
The non vanishing of the bond distortions even for long enough
tubes according to the Harigaya's model seems to be in
contradiction with experimental results for (3,3) nanotubes. For
(10,10) nanotubes both models work free of problems.

\section*{References}

(1) Dresselhaus, M. S., Dresselhaus, G. and Eklund, P. C. In:
"Science of Fullerenes and Carbon Nanotubes", New York: Academic
Press, 1996, p. 809.

(2) Ashcroft, N. W. and Mermin, N. D., "Solid State Physics", CBS
Publishing, Philadelphia, 1988.

(3) Su, W. P., Schrieffer, J. R. and Heeger, A. J., Phys. Rev. B
{\bf 22} (1980) 2099.

(4) Harigaya, K., J. Phys. Soc. Jpn. {\bf 60} (1991) 4001.

(5) Harigaya, K., Phys. Rev. B {\bf 45} (1992) 12071.

(6) Cabria, I., Mintmire, J. W. and White, C. T., Int. J. Quant.
Chem. {\bf 91} (2003) 51.

(7) Sünel, N. and Özsoy, O., Int. J. Quant. Chem. {\bf 100} (2004)
231.

(8) Özsoy, O. and Sünel, N., Czech J. Phys., {\bf 54} (2004) 1495.

(9) Qin, L. C., Zhao Xi, L., Hirahara, K., Miyamoto, Y., Ando, Y.
and Iijima, S., Nature 2000;408(6808):50.

(10) Sun, L. F., Xie, S. S., Liu, W., Zhou, W. Y., Liu, Z. Q.,
Tang, D. S., Wang, G. and Qian, L. X., Nature 2000;403(6768):384.

(11) Li, G. D., Tang, Z. K., Wang, N. and Chen, J. S., Carbon {\bf
40} (2002) 917.

(12) Stafstr\"{o}m, S., Riklund, R. and Chao, K. A., Phys. Rev. B {\bf
26} (1982) 4691.

(13) Ono, Y. and Hamano, T., J. Phys. Soc. Jpn., {\bf 69} (2000)
1769.

(14) Peng, L. M., Zhang, Z. L., Xue, Z. Q., Wu, Q. D., Gu, Z. N.
and Pettifor, D. G., Phys. Rev. Lett., {\bf 15} (2000) 3249.

(15) Harigaya, K. and Fujita, M., Phys. Rev. B {\bf 47} (1993) 16563.

\end{document}